\documentclass[twocolumn,floatfix,superscriptaddress]{revtex4-2}
\setcitestyle{super}

\usepackage{xfrac}
\usepackage{bm}

\usepackage{upgreek}

\usepackage[normalem]{ulem}

\usepackage{fancyhdr}
\usepackage{amsmath}
\usepackage{graphicx}
\usepackage{color}

\usepackage{lettrine}

\usepackage[unicode=true,bookmarks=true,pdfborder={0 0 0},backref=true,colorlinks=true,allcolors=blue,anchorcolor=black]{hyperref}
\usepackage{breakurl}

\makeatletter

\@ifundefined{textcolor}{}
{
 \definecolor{BLACK}{gray}{0}
 \definecolor{WHITE}{gray}{1}
 \definecolor{RED}{rgb}{0.7,0,0}
 \definecolor{ORANGE}{rgb}{1,0.25,0}
 \definecolor{GREEN}{rgb}{0,1,0}
 \definecolor{BLUE}{rgb}{0,0,1}
 \definecolor{CYAN}{cmyk}{1,0,0,0}
 \definecolor{MAGENTA}{cmyk}{0,1,0,0}
 \definecolor{YELLOW}{cmyk}{0,0,1,0}
}

\newcommand{\fpl}{\textcolor{black}}
\newcommand{\dom}{\textcolor{black}}
\newcommand{\loren}{\textcolor{black}}
\newcommand{\lmd}{\textcolor{black}}
\newcommand{\vor}{\textcolor{black}}
\newcommand{\rev}{\textcolor{black}}

\newcommand{\nanotec}{
CNR NANOTEC, Istituto di Nanotecnologia, Via Monteroni, 73100 Lecce, Italy
}

\newcommand{\unisal}{
Dipartimento di Matematica e Fisica E. de Giorgi, Universit\'{a} Del Salento, Campus Ecotekne, via Monteroni, Lecce, 73100, Italy
}

\newcommand{\mars}{
Aix Marseille Universit\'{e}, CNRS, Centrale Marseille, LMA UMR 7031, Marseille, France
}

\newcommand{\wolver}{
Faculty of Science and Engineering, University of Wolverhampton, Wulfruna Street, WV1 1LY, UK
}

\newcommand{\moscow}{
National Research Nuclear University MEPhI (Moscow Engineering Physics Institute), 115409 Moscow, Russia
}

\newcommand{\rqc}{
Russian Quantum Center, Skolkovo innovation city, 121205 Moscow, Russia
}

\newcommand{\iran}{
Department of Physics, Azarbaijan Shahid Madani University, Tabriz, Iran
}

\newcommand{\infn}{
INFN sezione di Lecce, 73100 Lecce, Italy\\
}

\renewcommand\frontmatter@abstractwidth{\dimexpr\textwidth-0.5in\relax}
\makeatother

\begin{document}

\title{
{
\usefont{OT1}{cmss}{m}{n}
{\Large
\textbf{Shaping the topology of light with a moving Rabi-oscillating vortex\\ 
}
}
}
}

\author{Lorenzo Dominici}
\email{lorenzo.dominici@nanotec.cnr.it}
\homepage[]{https://polaritonics.nanotec.cnr.it/}
\affiliation{\nanotec}

\author{Nina Voronova}
\email{nsvoronova@mephi.ru}
\thanks{cite this work as:  
\href{https://doi.org/10.1364/OE.438035}{Opt. Express \textbf{29}, 37262 (2021)}
}

\affiliation{\moscow}
\affiliation{\rqc}

\author{David Colas}
\affiliation{\mars}

\author{Antonio Gianfrate}
\affiliation{\nanotec}

\author{Amir Rahmani}
\affiliation{\iran}

\author{Vincenzo Ardizzone}
\affiliation{\nanotec}

\author{Dario Ballarini}
\affiliation{\nanotec}

\author{Milena De Giorgi}
\affiliation{\nanotec}

\author{Giuseppe Gigli}
\affiliation{\nanotec}
\affiliation{\unisal}

\author{Fabrice P.~Laussy}
\affiliation{\rqc}
\affiliation{\wolver}

\author{Daniele Sanvitto}
\affiliation{\nanotec}
\affiliation{\infn}

\begin{abstract}
\noindent \textbf{Abstract. }
Quantum vortices are the analogue of classical vortices in optics, Bose-Einstein condensates, superfluids and superconductors, where they provide the elementary mode of rotation and orbital angular momentum.
While they mediate important pair interactions and phase transitions in nonlinear fluids, their linear dynamics is useful for the shaping of complex light, as well as for topological entities in multi-component systems, such as full Bloch beams. Here, setting a quantum vortex into directional motion in an open-dissipative fluid of microcavity polaritons,
we observe the self-splitting of the packet, 
leading to the trembling movement of its center of mass, whereas the vortex core undergoes ultrafast spiraling along diverging and converging circles, in a sub-picosecond precessing fashion.
This singular dynamics is accompanied by vortex-antivortex pairs creation and annihilation, and a periodically changing topological charge.
The spiraling and branching mechanics represent a direct manifestation of the underlying Bloch pseudospin space, whose mapping is shown to be rotating and splitting itself. Its reshaping is due to three simultaneous drives along  the distinct directions of momentum and complex frequency, by means of the differential group velocities, Rabi frequency and dissipation rates, which are natural assets in coupled fields such as polaritons.
This state, displaying 
linear momentum 
dressed with oscillating angular momentum,
confirms the richness of multi-component and open quantum fluids and 
their innate potentiality to implement sophisticated and dynamical topological textures of light.
\end{abstract}

\maketitle

\usefont{OT1}{ppl}{m}{n}



\noindent {\large \textbf{1. Introduction}}\\
In quantum fluids and optical fields,
vortex cores are point-like phase singularities, a null-density \vor{point} around which the wavefunction
\vor{profile} rotates. Such zero-dimensional objects can become an axis of rotation extended along one direction inside a bulk space  
or when tracked in time \vor{on a plane}. 
This way, they draw curves 
inside a multidimensional domain ({\it e.g.}, of real space and time, but also in the reciprocal spaces of momentum and energy). These 
trajectories are called wave dislocations~\cite{nye_dislocations_1974}, or vortex lines and, being quantized, topological strings~\cite{volovik_universe_2009, Zurek1985}. Interesting examples are given by the extended networks of vortex lines inside a 3D atomic Bose-Einstein condensate (BEC), whose configurations and reconnections give rise to different phase transitions~\cite{Bewley2008, henn_emergence_2009, kondaurova_structure_2014}, analogously to what happens with electrons   
in superconductors, or to elementary pair interactions~\cite{dominici_interactions_2018} able to subtend diverse macroscopic patterns~\cite{zhao_pattern_2017}. Closed loops are possible, known as vortex rings~\cite{rayfield_quantized_1964, oholleran_topology_2006}, which can even become twisted to the extent of making vortex knots~\cite{Leach2005, Dennis2010, Kleckner2013, Kleckner2016}, as those induced in nonlinear optical media~\cite{Desyatnikov2012}. 
In optics, where angular \vor{momentum} quantization is pursued as a \fpl{possible} variable for data encoding~\cite{nagali_optimal_2009, leach_quantum_2010, fickler_quantum_2016}, optical vortex beams---like Laguerre-Gauss (LG), Bessel, and Airy beams with nonzero angular momentum~\cite{willner_different_2012, franke-arnold_advances_2008}---are similar possible solutions for 
propagation \vor{in empty space}. This confers robustness to free-space communications~\cite{krenn_twisted_2016, paterson_atmospheric_2005} and finds a variety of specialized applications, such as in femtochemistry, 
\vor{for} new selection-rules stud\vor{ies}, 
or \vor{for} 
ionized states manipulation in photoionization processes~\cite{schmiegelow_transfer_2016, picon_photoionization_2010}. 
\loren{
Also, advanced schemes 
for structured light 
have been proposed  
based on
frozen waves~\cite{dorrah_controlling_2016,zamboni-rached_carving_2021},
curved and solenoidal beams~\cite{Rahman2015,Zhao2015,Lee2010} with or without inner singularity, in order to realize trapping, rotational and pulling actions on small particles or objects, enriching the field of optical tweezers~\cite{Dholakia2002, nieminen_physics_2007, padgett_tweezers_2011, dholakia_shaping_2011}. 
Singular beams thus 
provide a great potential to study
topological complexity~\cite{soskin_singular_2017} and sophisticated applications, {\it e.g.}, in data encoding or interferometry gyroscopes.}

\fpl{In vectorial fields, a full description needs to be equipped with a vector charge or two dimensional-phase---prominently, transverse polarization in optics or pseudospin in multicomponent condensates---going beyond the scalar amplitude and phase of a scalar quantum fluid.}
The vortex core in a vector beam is not necessarily associated to a zero of the total density anymore~\cite{nye_wave_1987,berry_geometry_2001}, but it becomes a polarization singularity~\cite{dennis_chapter_2009,Cardano2013}, and it can represent a winding around any arbitrary axis of the 2D-valued phase~\cite{donati_twist_2016}, described by means of the Poincar\'{e} sphere in the case of a polarization state. 
This is what happens in so-called spin-vortices, {\it e.g.}, hedgehog and hyperbolic textures~\cite{Manni2013}, or lemon and star patterns in half-integer vortices~\cite{berry_geometry_2001, Cardano2013,Rubo2007}, and other generalized vortices~\cite{donati_twist_2016,Liu2015}.
\dom{The vector beam can be mapped to a 0D, 1D or 2D
domain \cite{lopez-mago_overall_2019} on the sphere, representing the set of carried pseudospin states and its codimension.
One of the most notable entities in the latter class is} the full Poincar\'{e} beam~\cite{beckley_full_2010}, an optical analogy of half-vortices~\cite{Rubo2007} and skyrmions~\cite{donati_twist_2016} in 
\vor{BECs, a sophisticated vector state that} has been shown to be generated simply by the 
\vor{overlap} of counter-polarized Laguerre-Gaussian states with a zero \vor{($\text{LG}_{00}$)} and unitary \vor{($\text{LG}_{01}$)} vortex charges. More recently, 
\vor{such a double-beam superposition state} has been 
\vor{enriched} with 
\vor{temporal} dynamics thanks to the Rabi \vor{oscillations}~\cite{Dominici2014,colas_polarization_2015} arising in strongly-coupled systems, such as polaritons \vor{that are themselves a superposition} 
of \vor{microcavity} photons and \vor{quantum well} excitons. 
This led to the realization of the so-called full Bloch beam~\cite{dominici_full-bloch_2021} (FBB) with a whole pseudospin \vor{(Bloch vector)} texture in the transverse plane which is also rotating and drifting in time,
\dom{
similarly to what has been theorized for rotating full Poincar\'{e} beams~\cite{krasnoshchekov_rotating_2017} and their polarization texture. 
As a consequence, using the natural filtering of the full-wavefunction offered by the cavity photoluminescence, rather than using a polarization filtering, the
Bloch singularity  was directly projected into the helicoidal dynamics \vor{of a vortex} line} sculpted into the light emission from the system.
\vor{Full Bloch beams} hosting Rabi\dom{-spiraling} vortices are also 
\vor{endowed} with 
time-varying orbital angular momentum (OAM), 
\vor{similarly} to those recently generated through nonlinear phenomena~\cite{ma_spiraling_2020,rego_generation_2019}. 
One way to understand such \vor{structured} emission is in terms of the mechanics found by Nye and Berry~\cite{nye_dislocations_1974, berry_elliptic_1979}, who showed that photonic helical dislocations in 3D space can be ascribed to the superposition of straight vortex lines with, {\it e.g.}, a tilted plane wave, and to their continuously varying relative phase, due to different wavevector directions. 
In the \vor{case of exciton-polariton} full Bloch beams, the relative phase is that between the upper and lower polariton modes, 
\dom{which is geometrically set by a two-pulse excitation scheme and}
\vor{evolving} in time due to the frequency difference induced by the Rabi 
\vor{splitting}. The \vor{resulting} emitted complex light is characterized by a spiraling inner vortex tube precessing around the axis of propagation, like a gyroscope, on a sub-picosecond timescale, or an ultrafast microscopic tornado of light whose axis of rotation is spiraling itself.\\

\noindent {\large \textbf{2. Results}}\\
\noindent \textbf{2.1 Polaritonic platform of coupled fields.}
Making use of the same polaritonic quantum-fluid platform, here we investigate \fpl{the \dom{rich} morphology of singularities \dom{achievable} in polaritonic fields. In particular, we show that} by 
\dom{combining} the transverse linear and orbital momentum of the fluid, 
\dom{we are able to unfold the pseudospin texture of the system, further empowering the dynamics in the observed photonic component}. 
\dom{Experimentally,} the present study relies on a single-pulse resonant excitation 
\vor{that imprints a linearly moving} vortex 
with the ultrafast tracking of its \vor{subsequent motion}.
While \vor{it is the internal} polariton \vor{Rabi oscillations that} drive the 
\vor{dynamics, the detection is being made} \vor{via} the external \vor{(emitted from the microcavity)} light field, and is therefore ultimately a pure singular-optics effect. This is a convenient, effective and powerful way to shape light. With linear motion, polariton composite vortices~\cite{Voronova2012} reveal additional non-trivial, counter-intuitive and peculiar features. 
The moving vortex is seen to start a self-oscillatory dynamics of its
core and center of mass, initially spiraling out and then
\vor{coming} back to the center of the packet. 
More sophisticated features such as ultrafast cycl\vor{ic}  
\vor{vortex-antivortex} pair-generation and annihilation events are observed, 
being the manifestation of the \rev{underlying Bloch pseudospin texture}.
\rev{It is important to note that while the spiraling of a vortex core as well as the creation of vortex-antivortex pairs are often observed and discussed in the context of nonlinear phenomena in optics, condensates and quantum fluids~\cite{dominici_vortex_2015}, here they are due to purely linear mechanisms. In essence, the strong coupling of the photon and exciton fields leads to new normal modes of the system, the polaritons, which have two different frequencies, and these two modes contribute to building up the observable photon field and its exciton counterpart. The dynamics in the resulting fields can then be described in terms of interference concepts, when taking into account a spatially varying relative phase between the normal modes and their continuous phase shift in time due to the different energies.} 
\vor{As a result,} the OAM 
of the observed photonic component is oscillating, \dom{sweeping several multiples of $\hbar$,} while \vor{staying con}served in the 
\vor{full-wavefunction} state comprising the (not directly observable) exciton \vor{component}. 
Despite the \dom{FBB} homeomorphism, consisting in a stereographic projection between the Bloch sphere and the real space, being here broken, the mapping of the Bloch parameters to real space is still possible and shown to \dom{mathematically subtend} all the dynamics.
This work highlights the complexity of vortex strings and topological surfaces obtainable with polariton fluids when driven by the internal dynamics of a full wavefunction. \dom{We used fundamental $\text{LG}$-beams building blocks to realize dynamical concepts of 2D pseudospin textures and their 1D and 0D connections.
The texture is realized and put into motion by the interplay of three driving effects, typical for polariton fluids, represented by the different group velocities, the  differential decay and the Rabi oscillations.} 
Future directions to explore include the implementation of such topology shaping to different platforms \cite{grillo_holographic_2015} or to even more variegated optical beams, to realise, {\it e.g.}, rotating vortex lattices
\dom{or advanced optical tweezers with rotational feature on multiple levels, {\it i.e.}, spin and orbital angular momentum, and spiraling vortex cores}.

\begin{figure}[htbp]
  \centering \includegraphics[width=8.4cm]{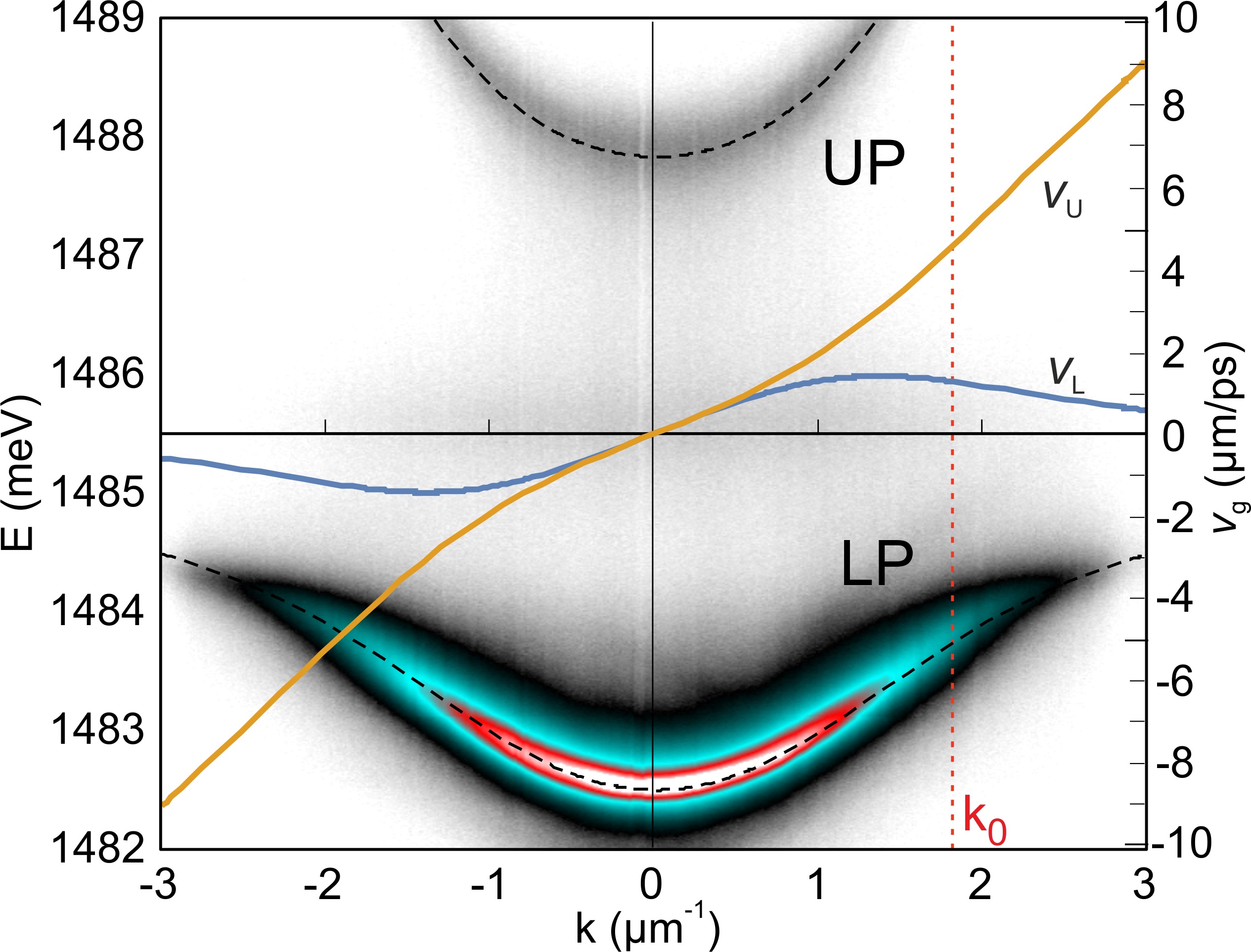}
  \linespread{1.0} \protect\protect
  \caption{\textbf{Polariton modes and group velocities}. Photonic emission from the polariton device under off-resonant excitation, highlighting the energy-momentum dispersion of the two normal modes, the upper and lower polariton branches. The dashed lines indicate the theoretical fitting of the dispersions $E(k)$. The modes splitting of 5.4~meV (at $k = 0$) corresponds to the Rabi period of $\approx 0.78$~ps. The blue and yellow solid lines show the group velocities $v_g = (1/\hbar)\partial E(k)/\partial k$ of the LP and UP packets motion, respectively.
       }
\label{FIG_branches}
\end{figure}
\begin{figure*}[htbp]
  \centering \includegraphics[width=12.8cm]{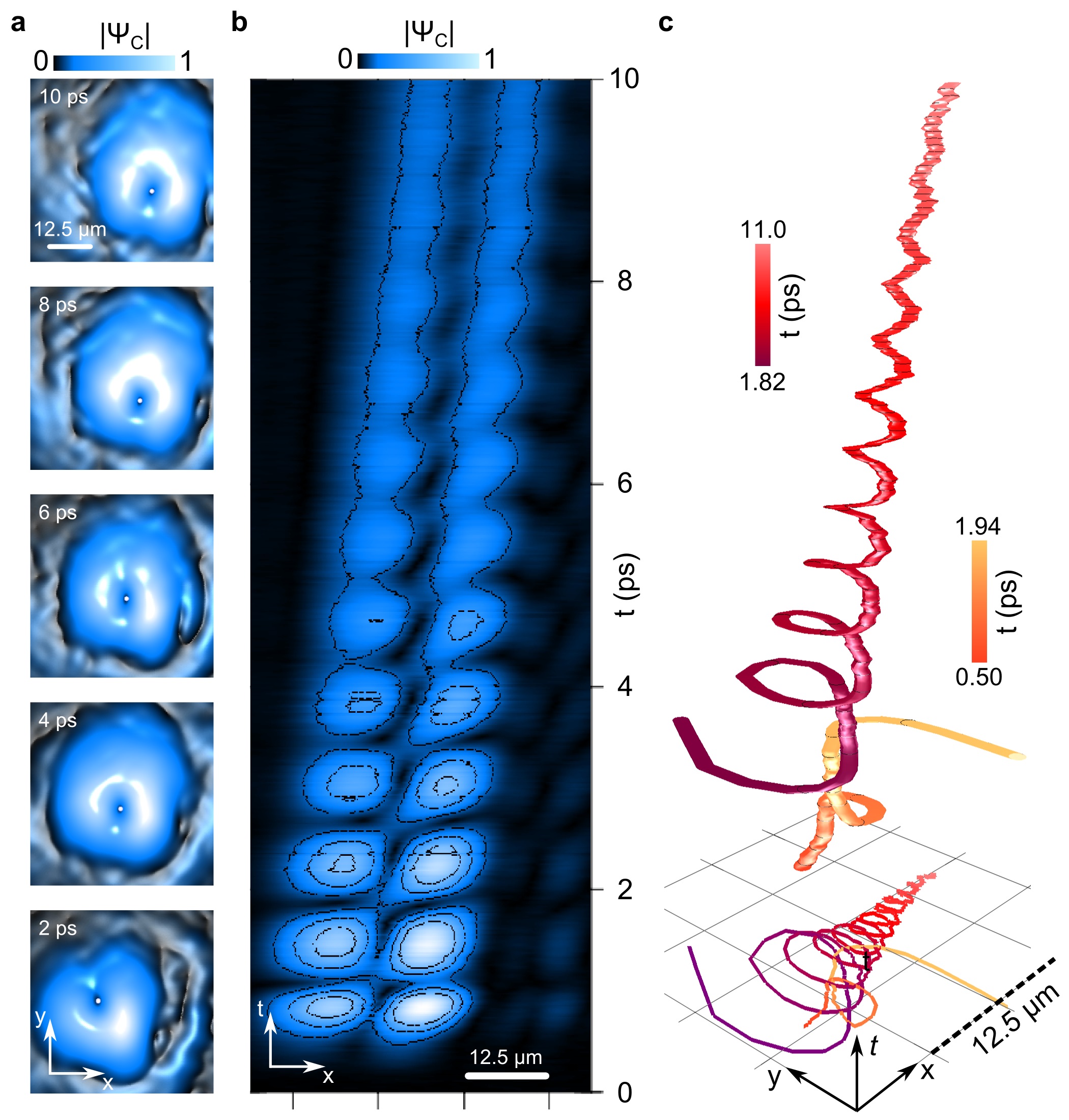}
  \linespread{1.0} \protect\protect
  \caption{\textbf{\vor{Full-range dynamics of the} moving Rabi-oscillating vortex.}
    \textbf{a}, The $xy$ space maps of the photonic amplitude \vor{$|\psi_C|$}
    at $t=2, 4, 6, 8 \text{ and } 10\text{ ps}$, as marked. 
    \textbf{b}, Timespace chart of the photonic amplitude taken along the $x$ direction of mo\vor{tion}, 
    displaying the Rabi oscillations and the spatial propagation of the vortex.
    \textbf{c}, Experimental vortex $xyt$ line (time range $t=0.5-11\text{ ps}$, step $\delta t=0.02\text{ ps}$). The trajectory is broken\vor{:} 
    in the range $t=1.82-1.94\text{ ps}$, two 
    \vor{coexisting} vortices are seen, the initial\vor{ly imprinted} one \vor{being} pushed to the boundary (orange curve, $0.50-1.94\text{ ps}$) and the one replacing it drawn to the centre \vor{of the cloud} (red curve, $1.82-11.0\text{ ps}$). The 
    \vor{latter vortex} bears the same winding and replaces the original one, but 
    \vor{rotates} in the opposite direction. The projection on the $xy$--plane illustrates the damped cycloid of the phase singularity \vor{position} during the vortex \vor{motion}.
    The launching momentum corresponds to $k_0=1.9~\mu$m$^{-1}$, and the average group velocity of the whole packet is $v_g=1.3~\mu \text{m}~\text{ps}^{-1}$. See also Visualization 1.
    }
\label{FIG_exp_moving}
\end{figure*}
%

\noindent \textbf{2.2 Moving Rabi\vor{-oscillating} vortex.}
The group velocity of a wavepacket represents the velocity at which both the $density$ and $phase$ spatial patterns move. 
\vor{In the case of exciton-polaritons, which are intrinsically a two-component fluid of coherently mutually-transforming excitons and photons~\cite{kavokin_book17a, Byrnes2014}, the new normal modes of the system---the upper and lower polaritons (UP, LP)---possess a peculiar non-parabolic energy dispersions.} \vor{In particular,} while the LP and UP modes 
have the same group velocity $(1/\hbar) \partial E(k)/\partial k$ at small in-plane momenta $k$, the two velocities ($v_L,v_U$) become \vor{distinct} as the LP mode $v_L$ is maximum at the inflection point of the $E(k)$ dispersion curve (see Fig.~\ref{FIG_branches})~\cite{gianfrate_superluminal_2018,colas_self-interfering_2016}. 
\vor{To exploit this peculiarity}, we 
\vor{imprint} a single $\text{LG}_{01}$ wavepacket 
by a femtosecond pulse at the central in-plane 
\vor{wavevector} $k_0=1.9~\mu$m$^{-1}$, slightly above the inflection point of the LP dispersion,
upon oblique incidence 
excitation. 
\lmd{The spatial width of the packet is $\sigma \sim 10~\mu$m.}
The central energy of the laser pulse has been blue-detuned with respect to the LP branch, in order to remain resonant with both normal modes to trigger the Rabi oscillations.
As a consequence, the \vor{exciton-polariton wave}packet splits into two fluids travelling at different speeds, while at the same time carrying an intrinsic orbital angular momentum (iOAM, {\it i.e.}, a vortex charge). As the initially coincident UP and LP vortices \vor{start} to continuously drift apart, they trigger 
\vor{the} dynamics 
\vor{in which the propagating vortex} cores in the photon and exciton components \vor{oscillate, displaying} rotati\vor{on} of both their center-of-mass and the phase singularity positions. 
Before \vor{turning} to the more complex early-time dynamics, we first show \vor{in Fig.~\ref{FIG_exp_moving}} the full range 
\vor{({\it i.e.}, long-time) dynamics} of the \vor{moving Rabi-oscillating} vortex. 
The \vor{wave}packet \vor{carrying a vortex} can be seen propagating, 
preserving its shape, as shown in the amplitude maps of Fig.~\ref{FIG_exp_moving}a, 
taken at successive time delays of 2~ps. The Rabi oscillations \vor{together with} the spatial propagation can be seen in Fig.~\ref{FIG_exp_moving}b, reporting the timespace chart of the photonic amplitude cut along the horizontal line of motion. \vor{The group velocity of the packet, extracted from t}he slope in such charts 
$v_g=1.3~\mu$m~ps$^{-1}$ \vor{is} in good agreement with the fitting of the LP mode dispersion, \vor{as the LP} 
branch is the one contributing most to the 
\vor{emitted} photon 
density at long times \vor{(}$\geq 2 \text{ps}$\vor{)}, due to the faster decay of the UP mode.
The full $xyt$ vortex line 
\vor{is presented} in Fig.~\ref{FIG_exp_moving}c. At 
\vor{early} times ($\leq 2 \text{ps}$, orange lines), the initial\vor{ly imprinted} vortex core 
gradually starts a (clockwise) rotational motion with increasing radius, \vor{corresponding to} an increasing in-plane speed.
The vortex line is broken as soon as the core gets displaced outside the density
\vor{spot}, during the second Rabi period, and it is replaced by a co-winding but oppositely (counter-clockwise) spiraling vortex drawn from outside \vor{of the packet} ($\geq 2 \text{ps}$, red lines).
The 
\vor{rotation direction} inversion is ascribed to the inversion of the \vor{overall} UP and LP populations \vor{imbalance}. By tuning the laser \vor{excitation} energy, 
the initial state \vor{was brought} closer to the UP mode, \vor{{\it i.e.}, at early times, the system is characterized by}
a larger content of upper polaritons. \vor{Since they are} 
decaying faster \vor{than the LP,} at a given time, they reach equal 
\vor{populations}.
After the population reversal, \vor{the lower polaritons always dominate in the system, and} the phase singularity is tracked as a continuous string. 
Its structure consists of a translational motion 
\vor{combined with the} Rabi rotation, 
\vor{and the trajectory} is 
a damped cycloid on the $xy$--plane. The linear movement is also associated to the whole vortex packet,
with the core which would reside in its centre during \vor{the propagation} 
(in the absence of oscillations), while the Rabi-induced rotation periodically brings the vortex core inside and outside the centre.\\
%

\begin{figure}[htbp]
  \centering \includegraphics[width=8cm]{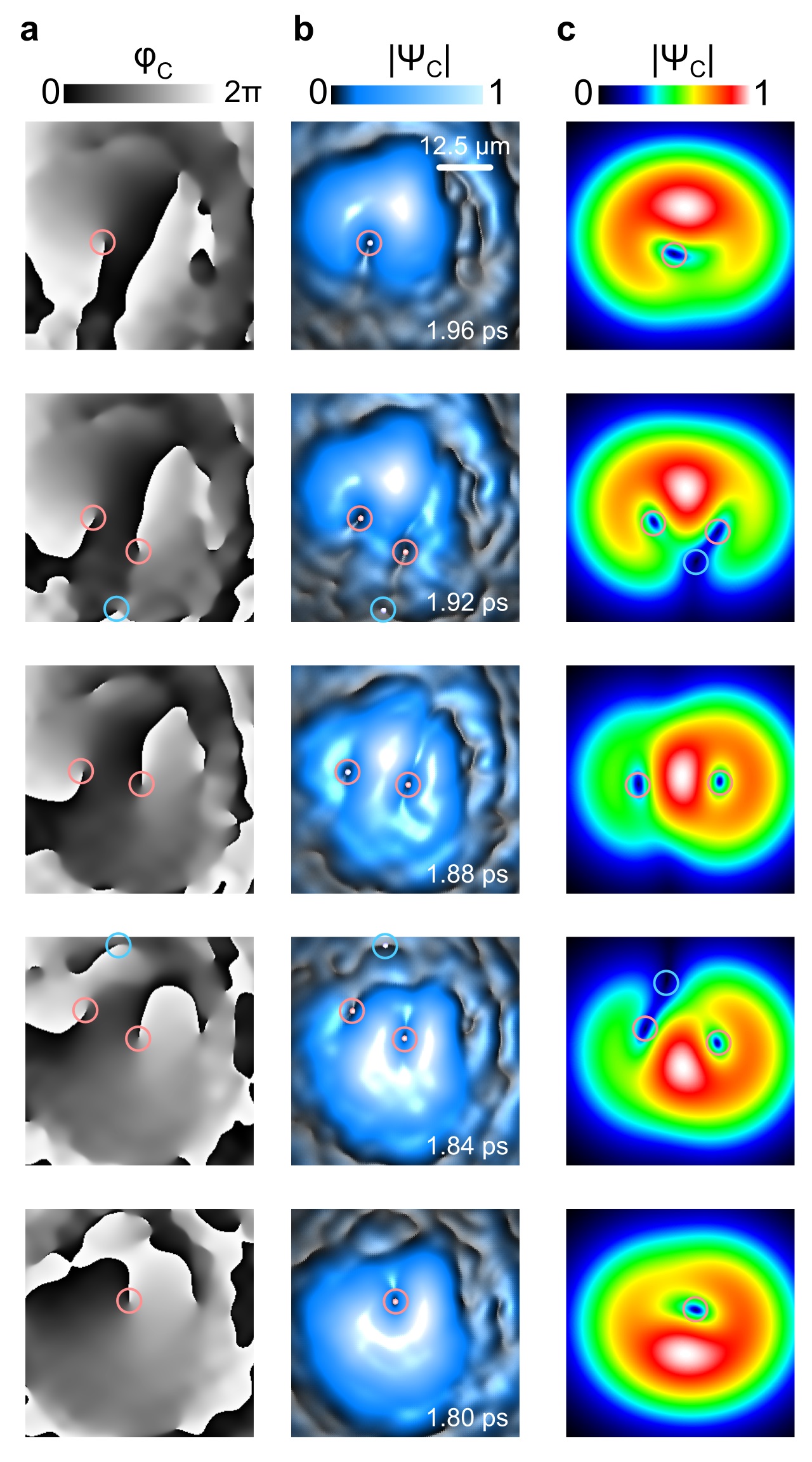}
  \linespread{1.0} \protect\protect
  \caption{\textbf{Early-time dynamics.}
    \textbf{a, b}, Experimental space maps of the photonic phase \vor{$\varphi_C$} (a) and amplitude \vor{$|\psi_C|$} (b) of the propagating vortex
    \vor{shortly} after the excitation (bottom to top, isospaced by 40 fs). The initial phase singularity is 
    moved out and replaced by a core of the same charge, created close to the boundary in a \vor{V--AV} pair creation process. The opposite\vor{ly} charge\vor{d vortices} which are created \vor{together} with the new, and \vor{then} annihilated with the original 
    \vor{vortex} are visible in the \vor{panels corresponding to} $t = 1.84~\text{ps}$ and $1.92~\text{ps}$, respectively.
    \textbf{c}, The 
    \vor{moving LG beams} model reproducing the initial 
    \vor{vortex} swapping by the pair-creation and -annihilation.
    \vor{In all panels, the vortex and antivortex cores positions are marked with the pink and blue circles, respectively.}
    See also Visualization 1.
    }
\label{FIG_early_times}
\end{figure}

\noindent \textbf{2.3 Vortex-antivortex pair creation and annihilation.} 
\vor{Whereas the combination of translational and Rabi-oscillatory rotational motion explains the continuous spiraling trajectory (the red line in Fig.~\ref{FIG_exp_moving}c), the discontinuity of the vortex line requires} a more careful look for the early-times dynamics \vor{which is presented in Fig.~\ref{FIG_early_times}}. 
\vor{In particular,} the 
\vor{imprinted} vortex core 
(see the bottom panel corresponding to $t=1.80$~ps)
and 
a further 
vortex \vor{entering the spot} 
from the outside, 
\vor{can be simultaneously} visible in
\dom{the successive time frames of} Fig.~\ref{FIG_early_times}a,b (amplitude and phase of the photonic emission 
at 40~fs equispaced time frames, taken around \vor{$t\approx2$~ps}).

%
This dynamics
\rev{suggests that the vortex initially seen is associated to the UP wavepacket, while after its faster decay, the vortex is associated to the LP mode. In the intermediate times, two vortices with the same sign are visible. The density map does not allow to precisely track where the excess vortex is coming from. However the phase maps suggest that the pair creation events at the boundaries play a role in the disappearance of the initial vortex and the appearance of the new one. Such events are shadowed by the sensitivity to noise of the external region of the polariton fluid bearing very low density. However, the theoretical model confirms the mechanics,}
which is first accompanied by the appearance of a vortex-antivortex (V--AV) pair, with the AV that quickly escapes the condensate spot (see the panels corresponding to $t=1.84$~ps)
leading to the 
\vor{simultaneous} presence of the two \dom{Vs} 
\vor{of the same charge} 
\dom{which are neatly visible in the middle of the packet ($t=1.88$~ps), and}
bringing the iOAM of the photon fluid to $2\hbar$. 
\rev{As shown later, the mechanism is in agreement with the presence of one open edge line able to move the AV to and from the transverse infinity and hence produce the dynamic variation
of the net topological charge in the observable region of space~\cite{molina-terriza_vortex_2001}.}
\dom{We anticipate here and discuss later that the total OAM is conserved in the global wavefunction made of both photon and exciton fields.} Soon after, the AV reappears from the opposite side of the sample ($t=1.92$~ps), annihilating with the initially imprinted vortex and restoring the unitary topological charge of the fluid. The remaining vortex ($t=1.96$~ps), created in the V--AV pair generation event, starts spiraling in the opposite direction, as described above.  
The \vor{panels in the} third column, Fig.~\ref{FIG_early_times}c, \vor{contain} a reconstruction of the photonic amplitude \vor{dynamics} by means of a \vor{model overlapping the two} polariton 
$\text{LG}_{01}$ 
\dom{packets of the same width but
with a different center and amplitude, and a shifting relative phase.
While the first two parameters are set by the packets
moving with different velocities and by their different emission rates, the geometrical relative phase cycles due to the Rabi energy splitting, as discussed in the following.\\}

\begin{figure}[htbp]
  \centering \includegraphics[width=8cm]{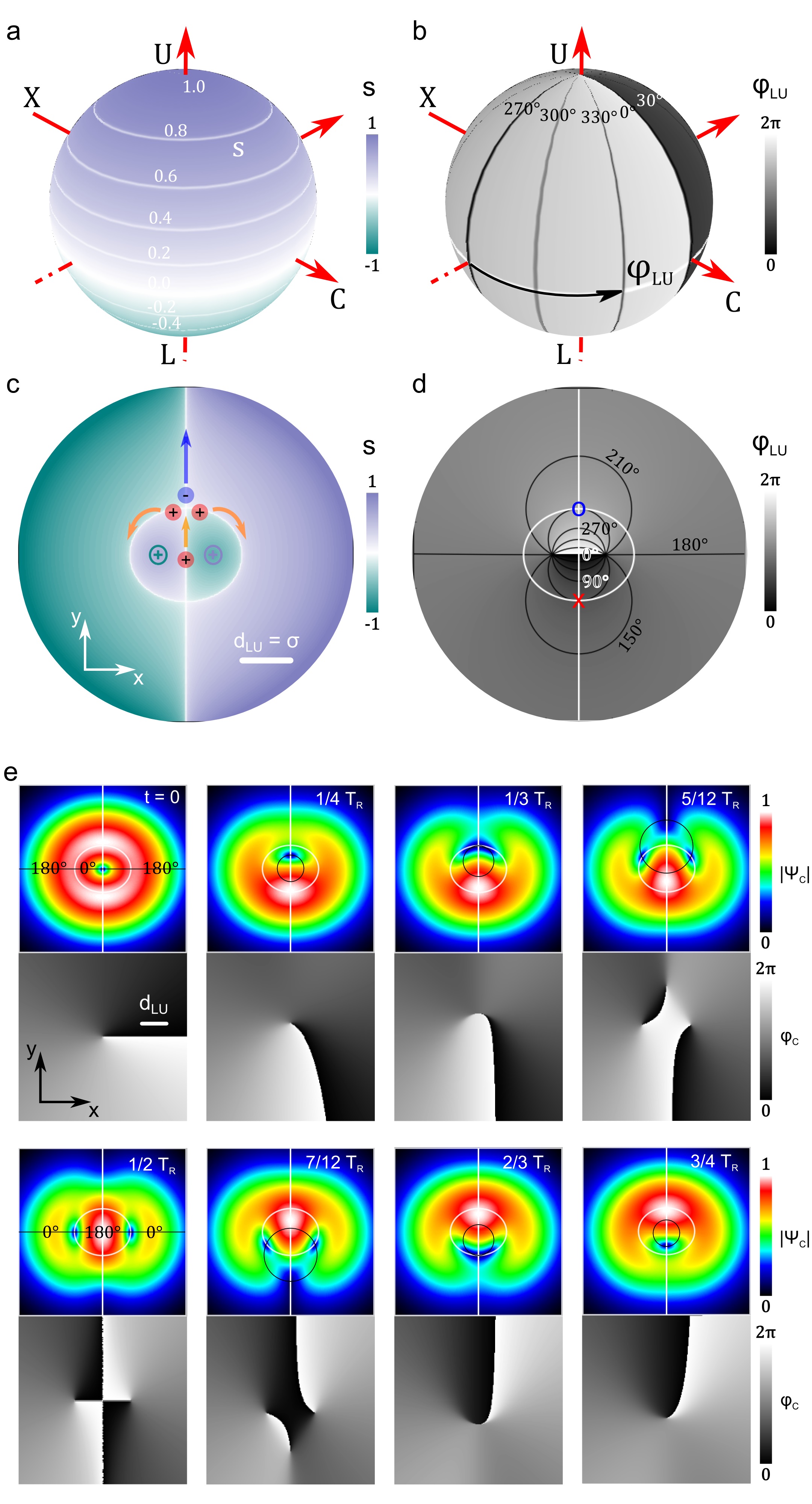}
  \linespread{1.0} \protect\protect
  \caption{\textbf{\vor{Bloch sphere to real plane mapping and the
  V-AV pair creation/annihilation} 
  during a Rabi cycle 
  in the symmetric content case $\bm{S=0}$.}
    \textbf{a,b}, \vor{Exciton-polariton} Bloch sphere,
    \vor{with its latitude} $s$ 
    (a) and  
    longitude $\varphi_{LU}$ (b) 
    \vor{coordinates and color-scale maps}.
    \textbf{c,d}, The model \vor{real-plane} maps of the \vor{local} content \vor{imbalance} 
    $s(x,y)$ and relative phase $\varphi_{LU}(x,y)$. 
    The intersecting vertical line and a 
    loop (solid white in {\bf c}, {\bf d} and {\bf e}) represent the loci of the isocontent $s=0$  
    \vor{which is the trajectory of} the exciton and photon vortices \vor{cores}. 
    \vor{In {\bf c}, the red and blue dots} represent the initial\vor{ly imprinted photonic V} 
    moving up and split\vor{ing} into two  
    Vs and an AV at the intersection point between the straight line and the loop.
    The LP and UP cores are marked with green and purple circles, respectively.
    \dom{
    In {\bf d}, the $\varphi_{LU}$ isolines are marked with solid black half-circles at every $30^\circ$ intervals (which is a $T_R/12$ time spacing).}
    \rev{The places for the $\text{V} \rightarrow 2\text{V}+\text{AV}$ and $2\text{V}+\text{AV} \rightarrow \text{V}$ events are shown with the open blue circle (creation) and the red cross (annihilation). 
    They are distant in time four isolines, $4/12~T_R = 1/3~T_R$}
    \textbf{e}, \vor{Maps of $|\psi_C|$ and $\varphi_C$} 
    at different times \vor{during one Rabi period}, 
    show\vor{ing} the initial photonic \vor{V} 
    moving up along the line and swap\vor{ping} into a charge \vor{triplet} (V plus 
    \vor{a V-AV} pair). 
    The two  
    \vor{Vs} move laterally and downwards along the circle, and the \vor{AV} 
    \vor{keeps moving} up \vor{until going to infinity and then reappearing} from the bottom of the straight line before annihilating with the two 
    \vor{Vs} in the last panel, ready to start the cycle again.
    \dom{The solid black circle represents the union of the two $(\varphi_{LU} = 0) \cup (\varphi_{LU} = \pi)$ meridians at each time.}
}
\label{FIG_isocontent_cases}
\end{figure}
%

\noindent \textbf{2.4 Polariton Bloch sphere mapping.}
We now describe the \vor{topological} texture 
underlying 
the dynamics observed in the photonic component \vor{of the polariton full wavefunction created on the exciton-polariton Bloch sphere  by the continuously shifting (non-coinciding) wavepackets in the two system's normal modes. The exciton component is} 
expected to be\vor{have similarly, being} 
in phase opposition 
\vor{to the photon state (see the Bloch sphere metric introduced in Fig.~\ref{FIG_isocontent_cases}}). 
In a perfect analogy to the Poincar\'{e} sphere for polarization, we use polar coordinates to map both \vor{the local UP-LP} content \vor{imbalance} 
and relative phase \vor{between the UP and LP fields}. 
\vor{Similarly to} the Stokes parameter $s_3$
~\cite{lopez-mago_dynamics_2013}, the content \vor{imbalance} parameter $s = \frac{|\psi_U|^2-|\psi_L|^2}{|\psi_U|^2+|\psi_L|^2} = \cos{\theta}$ ($\theta$ being the polar angle) defines the latitude on the sphere, while  
the longitude, or azimuthal angle, is directly the relative phase $\varphi_{LU}=\varphi_U-\varphi_L$ \vor{(assuming $\psi_{L(U)} = |\psi_{L(U)}|\exp[i\varphi_{L(U)}]$)}.
The two \vor{respective variables} are represented in Fig.~\ref{FIG_isocontent_cases}a,b 
both as a coordinate system and by color-scale 
\vor{distributions} on the sphere, \vor{where the metric is always fixed, while the evolution of the polariton system consists of the redistribution of the full wavefunction according to the Rabi rotation, packets motion, decay, {\it etc.} (for details, see Ref.~\cite{dominici_full-bloch_2021}).
At the same time, on the real plane, the respective parameters $s=s(x,y)$ and $\varphi_{LU}=\varphi_{LU}(x,y)$ are always shifting due to the change in the UP and LP wavefunctions at each point of space.}
We also define the global content \vor{imbalance} parameter $S=\frac{\int |\psi_U|^2d{\bf r}-\int|\psi_L|^2d{\bf r}}{\int|\psi_U|^2d{\bf r}+\int|\psi_L|^2d{\bf r}}$ that characterises the system and the current shape of $s$ mapping between the Bloch sphere and the real space at any given time.
\rev{The morphology of the $s$ and $\varphi_{LU}$ maps is fundamental to understand the vortex cores dynamics.
It can be stated that the vortex core in one field (here photon or exciton), being not just a phase singularity but also a density zero by definition, can only occur at a point where the normal modes (upper and lower polaritons) fully destructively interfere, {\it i.e.}, at the point where their amplitude is the same. For such a reason, the vortex core in the photon (or exciton) field can only move along a path which is represented by the $s(x,y)=0$ lines. The other condition required for destructive interference is the phase difference between the base fields being $\varphi_{LU}(x,y) = 0$ (or $\pi$) for the photon (or exciton) field. The relative phase whose shape is that of a vortex dipole is then rotating due to the Rabi oscillations. In other terms, the vortex core in the photon (or exciton) field move along the isocontent orbits, according to the cyclic Rabi-oscillations drive.\\}

\begin{figure}[htbp]
  \centering \includegraphics[width=9.0cm]{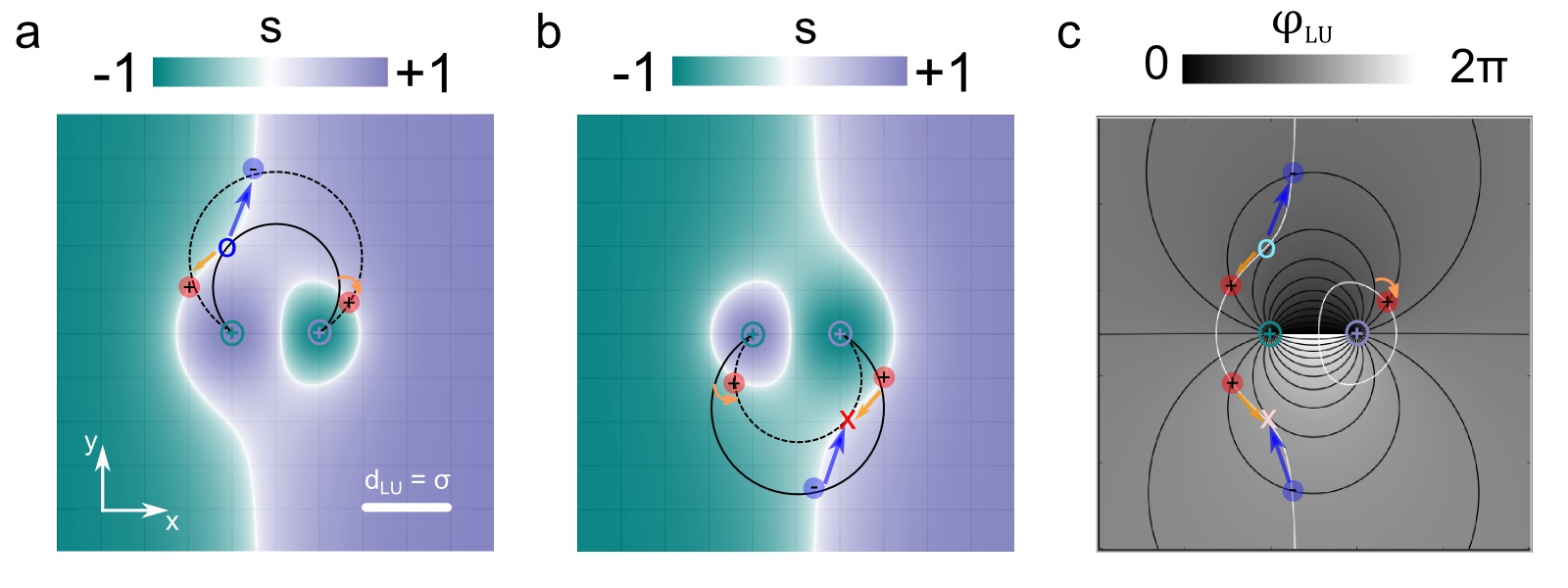}
  \linespread{1.0} \protect\protect
  \caption{\textbf{
    Photon vortex cores orbits and \vor{V-AV} pair dynamics 
    in the general 
    case $\bm{S\neq0}$.}
    \textbf{a,b}, $s(x,y)$ maps with $S>0$ \vor{(0.18, {\bf a})} and $S<0$ \vor{(-0.18, {\bf b})}. The main \vor{(originally imprinted) vortex} 
    is circulating along the closed $s=0$ loop (white loop). While the main \vor{vortex} charge maintains the same initial sign ({\it e.g.}, positive), its core moves in the opposite directions in the two \vor{respective} cases. The 
    \vor{V-AV} pair 
    \vor{creation and annihilation cyclically} happen on the open branch 
    of the 
    $s=0$ isocontent, 
    in the two different halves of the Rabi period, and are shown with the open blue circle \vor{(creation)} and the red cross \vor{(annihilation)} 
    on the maps. 
    \textbf{c}, \vor{$\varphi_{LU}(x,y)$ map showing} that these events happen in the spatial positions where the relative isophase lines (\vor{represented by the} solid black \vor{line}, with a $2\pi/24$ spacing) are tangent to the 
    \vor{vortex trajectory} (white solid line). 
    \dom{As an example, the solid and dashed black lines in \textbf{a,b} represent the circular arc of $\varphi_{LU}=0$, at a given time and after $T_R/24$, respectively}.
    \rev{The pair events in the case of a constant $S=0.18$ are shown in \textbf{c} as distant in time six isolines, equal to $6/24~T_R = 1/4~T_R$.}
    \lmd{The LP and UP cores are marked with green and purple circles, respectively.} 
    \rev{When considering together the Rabi cycle and the differential decay, the orbital motion is following the orbit reshaping, and the dynamics can evolve from panel \textbf{a} to panel \textbf{b}, with the main vortex swapping from the right to the left side during the population inversion ($S$ crossing zero), as in the experiments of Fig.~\ref{FIG_early_times}.}
    See also Visualization 2.
}
\label{FIG_asymmetric_case}
\end{figure}

\noindent \textbf{2.5 Symmetric content case.}
The $s$ and $\varphi_{LU}$ maps, theoretically retrieved, are presented in Fig.~\ref{FIG_isocontent_cases}c,d, respectively. 
At early times, the vortex cores \vor{in the overlapping UP and LP fluids coincide}, 
and $s$ is constant all over the 
\vor{real plane}, but 
\vor{with time, the} distance \vor{between the cores} linearly increases
as $d_{LU} = (v_U - v_L)t$.
The 
\vor{distribution} represented \vor{in  Fig.~\ref{FIG_isocontent_cases}c,d corresponds to} a given reached distance \rev{($d_{LU}=\sigma$)} between the vortices in the UP and LP components of the fluid and, for simplicity, for the case of equal global UP and LP contents ($S=0$).
It is important to point out a fundamental difference with the case of the full Bloch beam 
\vor{created} by a double-pulse splitting of \vor{two} standing vortices~\cite{dominici_full-bloch_2021}.
In the case \vor{of the FBB}, the metric of meridians and parallels \vor{on the sphere is} 
mapped to the real 
plane 
\vor{as} two families of orthogonal \vor{Apollonian} circles.
\vor{In the current case}, there is no second $\text{LG}_{00}$ pulse shaping 
the normal modes packets into asymmetric, off-axis vortices. \vor{Contrary to FBB, here} they remain two symmetric $\text{LG}_{01}$ vortices, but splitting their positions in time \vor{due to the difference in group velocity of the upper and lower polaritons}.  
On the one hand, this 
\vor{similarly creates} a vortex dipole, but with increasing size, in the relative phase $\varphi_{LU}$, \vor{as shown in} Fig.~\ref{FIG_isocontent_cases}d. \lmd{The isophase meridians do form a circle on the sphere when taken together in pairs, $(\varphi_{LU} = \text{const}) \cup (\varphi_{LU} = \text{const} + \pi)$, and these circles are still mapped into circles in real space passing 
\vor{through the positions of the UP and LP vortex} cores (see also Fig.~\ref{FIG_asymmetric_case}c).} 
On the other hand, the isocontent \vor{lines $s(x,y)=$~const, representing the trajectories of any specific quantum state in real space,} are not perfect circles anymore, and hence not everywhere \rev{orthogonal to the relative phase isolines $\varphi_{LU}(x,y)=$~const.} 
The pure photon (C) and exciton (X) are the \vor{phase-opposite} states on the equator of the Bloch sphere \vor{(see Fig.~\ref{FIG_isocontent_cases}a, the $s=0$ line)}, 
corresponding to the intersection of this line with \vor{the longitudes} $\varphi_{LU}=0$ and $\pi$ \vor{(see Fig.~\ref{FIG_isocontent_cases}b)}. The isocontent lines \vor{on the real plane}, $s(x,y)=0$, shown in white in the Fig.~\ref{FIG_isocontent_cases}c-e, map 
the trajectories followed by the 
\vor{C (X)} vortex cores which, 
being null-density points in the respective fields, are \vor{the only} points \vor{in space 
hosting their \lmd{pure} X (C)} counterpart. The Rabi \vor{oscillations, corresponding to the full-wavefunction rotation on the Bloch sphere,} are understood as 
\vor{a continuous drift of the isophase lines on the plane that 
allows to track 
the dynamics of the vortex cores 
moving along the $s(x,y)=0$ lines following the intersection with the isophases $\varphi_{LU}=\pi$ or 0}. 
Here, \vor{contrary to the case of the \lmd{standing} FBB state}, the shape and size of the vortex core orbits are \vor{not fixed, but} change in time 
even in the conservative case of no decay \vor{({\it i.e.}, when $S=$~const)}, \lmd{due to the relative movement}.\\

\noindent \textbf{2.6 Three-charges events.}
We first show for clarity the mechanics in the ideal case of both the absence of decay and of frozen packets, where the relative movement is suppressed (but after they already have split by some distance). 
As a consequence, the effective dynamics retrieved by the theory for the \vor{photonic amplitude} $|\psi_C|$ (\vor{or, similarly, the exciton amplitude} $|\psi_X|$), 
\vor{their} phase and vortex cores in the symmetric $s(x,y)$ situation is shown in Fig.~\ref{FIG_isocontent_cases}e at significant time frames during one Rabi-\vor{oscillation period} 
$T_R$.
\vor{To track the C (or X) vortex cores motion, one needs to follow the intersection of the $s(x,y)=0$ 
isocontent line\lmd{, representing the motion trajectories,}
with the isophase $\varphi_{LU}(x,y)=\pi$ (or 0) in time, 
\lmd{representing the motion drive}
induced 
by the Rabi rotation of the full wavefunction on the Bloch sphere}.
In the present case, \lmd{there are two nodal points of the} \lmd{isocontent line with itself ({\it i.e.}, such a line is composed by the intersecting loop and straight line), so that there are two points of trajectory bifurcation. As a result, these positions become}
singular points where the 
\vor{V-AV pairs} creation and annihilation events periodically happen.
\lmd{In fact, both V and AV in the photon field share the same property in the basis of the normal modes, being a zero of the respective field and lying on the same equatorial point of the Bloch sphere.} The consequence is that a vortex in the centre that is moving upward would split
\rev{(precisely at the time 
$t=1/3~T_R$)} 
 into two vortices going left and right 
and an anti-vortex that continues to go up. The  
AV is seen going to infinite distance ($y\to\infty$), \vor{coming back} 
at a successive time 
from the opposite direction ($y\to-\infty$), 
\vor{after which} the three \vor{vortex} charges annihilate
\rev{(at 
$t=2/3~T_R$, 
as also seen from the relevant isophase line in Fig.~\ref{FIG_isocontent_cases}d)}, 
\vor{leaving one} resulting 
vortex \vor{in the center}.
\rev{The creation and annihilation events are hence separated in time by $\Delta t = (2/3 - 1/3)~T_R = 1/3~T_R$, while the experimental sequence in Fig.~\ref{FIG_early_times} is comprised in a $0.16$~ps time interval, which is $1/5~T_R$. Such a difference can be ascribed to the varying separation $d_{LU}$ between the lower and upper vortex cores, which is fixed and equal to the packet width $\sigma$ in the model case, while it has only reached a lower value (roughly estimated to be less than $0.65~\sigma$) in the experiments. Such a distance influences the orbits shape, as discussed in the following section, and also affects the positions and time at which V-AV pair events take place.}
\loren{It is noteworthy that while there are always two vortices in the normal modes' relative phase $\varphi_{LU}$, and either one or three vortices at a time in a single bare mode ({\it i.e.}, photon or exciton), there are either two or four singularities in the relative phase between the bare modes [$\varphi_{CX}(x,y)$, not shown]. In such a phase map, the two orbital splitting points said above represent two saddle points~\cite{nye_phase_1988,berry_much_1998}, and the specific $(\varphi_{CX} = \pi/2) \cup (\varphi_{CX} = 3\pi/2)$ isoline concides with the $s=0$ isocontent line shown in Fig.~\ref{FIG_isocontent_cases}.}\\

\begin{figure}[htbp]
  \centering \includegraphics[width=9.1cm]{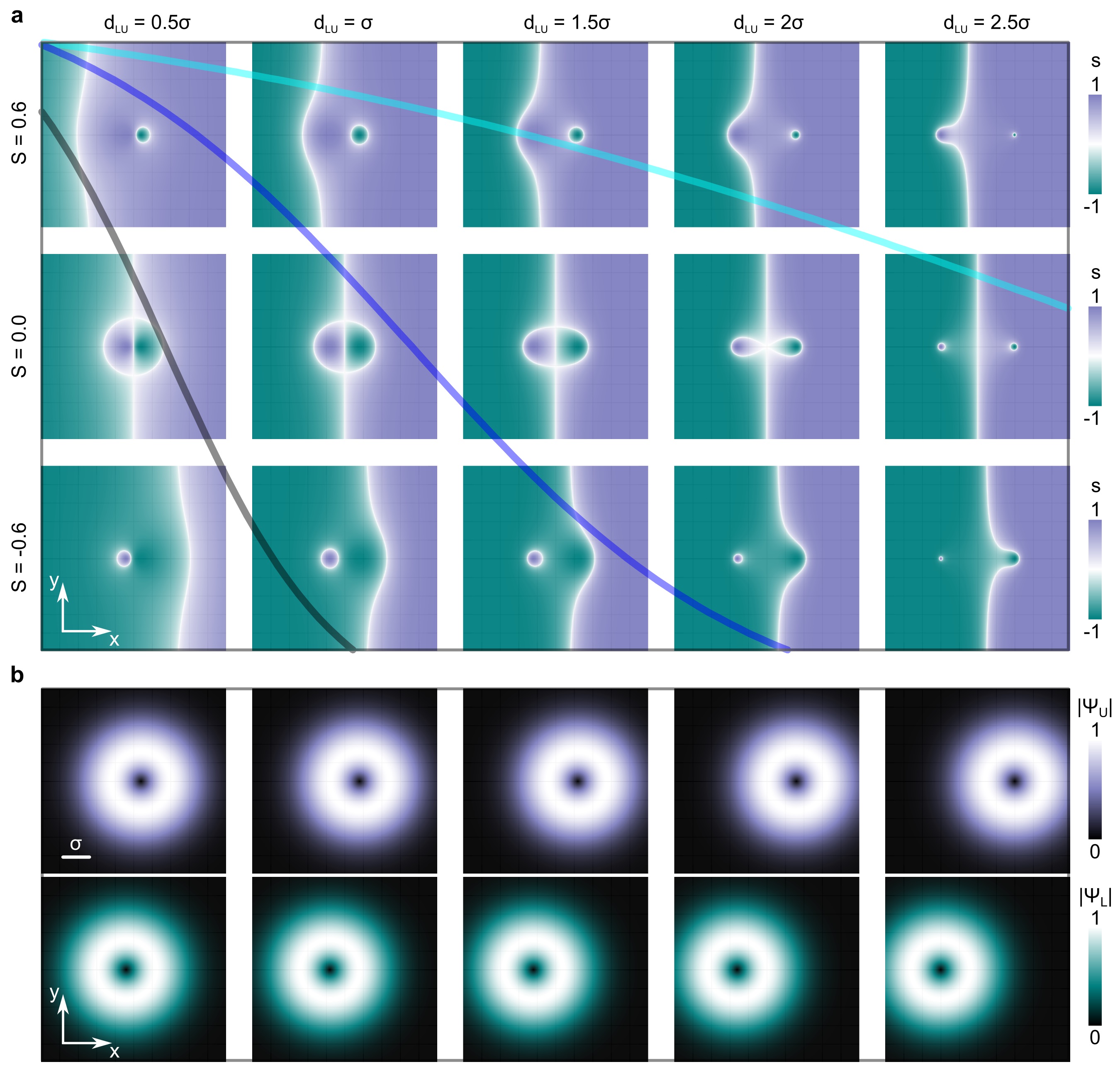}
  \linespread{1.0} \protect\protect
  \caption{
  \textbf{Local imbalance maps and polariton vortex packets.}
    \textbf{a}, \vor{Local imbalance real-space maps} $s(x,y)$ 
    \vor{for} different cores displacements $d_{LU}$ \vor{(columns)} and global 
   \vor{imbalances} $S$ (rows).
    The 
    \vor{C (X) vortex} cores move, \vor{driven by} 
    the Rabi 
    \vor{rotation of the polariton full wavefunction on the Bloch sphere,} along the 
    line $s(x,y)=0$ (white lines). 
    \vor{While the distance} $d_{LU}$ \vor{changes in time linearly, the evolution of the global imbalance parameter} 
    $S$ 
    \vor{assumes the hyperbolic tangent shape}. 
    \vor{Therefore, the} real dynamics \vor{effectively follows the maps 
    along} the overlapped solid lines (the specific 
    \vor{line depends} on the initial \vor{value of} $S$ \vor{defined by the excitation laser and on the} \lmd{$(v_U-v_L)/\sigma$ to $\gamma_{LU}$ ratio}). \textbf{b}, \vor{Schematics of} the moving LP and UP \vor{$\text{LG}_{01}$ wave}packets for different relative distances $d_{LU}$ (according to the same column ordering as in {\bf a}). 
    For the sake of representation, 
    \vor{panels in both {\bf a} and {\bf b} are centered} 
    with respect to the moving 
    \vor{middle of the vortex dipole} $x_0 = (v_U - v_L)t/2$ ({\it i.e.}, the \vor{UP and LP vortices} are always positioned symmetrically). 
    See Visualization 2.
}
\label{FIG_table_cases}
\end{figure}

\noindent \textbf{2.7 Relative decay and asymmetric dynamics.}
In the experiments, while the vortex dipole in the relative phase \vor{profile} expands due to the gradual \vor{drifting apart} 
of the LP and UP \vor{vortex} cores during \vor{their motion}, 
the \vor{local} \vor{imbalance} parameter \vor{$s(x,y)$} 
redistributes due to both the \vor{wavepackets} motion and the differential decay of the two modes, $\gamma_{LU}=\gamma_U-\gamma_L\approx \gamma_U$ \vor{(the decay rates are introduced as $|\psi_{L(U)}(t)| = \exp[-\gamma_{L(U)}t]|\psi_{L(U)}(0)|$)}.
\vor{The interplay of the two effects} is also reflected on the photon (and exciton) \vor{vortex} core orbits which initially gradually 
\vor{expand}, 
reach a maximum \vor{size} and then 
\vor{shrink} back.
Figure~\ref{FIG_table_cases}a 
shows the model maps of $s(x,y)$ for the case of larger, equal, and smaller total UP to LP content ($S>0$, $S=0$, and $S<0$, top, middle, and bottom row panels, respectively). Each column \vor{of the table} corresponds 
to increasing relative distances between the \vor{vortex} cores \vor{in the system's normal modes} (whose packets appear in Fig.~\ref{FIG_table_cases}b). The figure \vor{illustrates} 
symmetric and asymmetric \vor{local imbalance} maps corresponding to 
\vor{different} situations, 
in particular, to the evolution of the \vor{$s=0$} lines that 
\vor{are the trajectories of} the photon and exciton vortex cores motion (white lines in the maps, separating domains of $s>0$ and $s<0$). They are composed of a closed ring and a bent open line ({\it i.e.}, the line closed at infinite distance) that eventually \vor{intersect and} merge \lmd{(}
\vor{in some panels,} \lmd{they 
become two closed loops and one open line). 
Conditions for their intersection and detachments are described in a previous work on the positioning of polarization singularities~\cite{lopez-mago_dynamics_2013}.} In the case of the moving polariton vortex packets, their evolution is due to the \vor{linear increase of the} relative distance $d_{LU}$ 
in time and the global population imbalance $S$ decrease according to the exponential decay of the two polariton components, yielding $S(t) = \tanh\{-\gamma_{LU}t -\text{atanh}[S(0)]\}$. The 
\vor{actual} evolution of the $s(x,y)$ maps is hence 
\vor{the sequence of maps following the lines overlapping the table of frames in Fig.~\ref{FIG_table_cases}a} (cyan, blue and black solid lines), given by different $\gamma_{LU}$ or $S(0)$.

In the symmetric case of Fig.~\ref{FIG_isocontent_cases}, the cores dynamics comprises a three-charges event, which in reality is rarely observed 
as the \vor{balance of total populations, providing the symmetric case} $S=0$, 
is reached only 
\vor{once during the evolution and then is instantly} lost. 
The dynamics in a more realistic case of the asymmetric distribution (both $S>0$ and $S<0$) is shown in Fig.~\ref{FIG_asymmetric_case}a,b. While one 
\vor{initially imprinted (``main'') vortex} core is moving along a closed loop on one side of the 
\vor{polariton cloud}, the separated line\vor{, also mapping the Bloch sphere equator,} 
\vor{hosts the} two-charges events. Subsequent vortex-antivortex pair creation and annihilation 
(marked with the open circle and cross marks, respectively) are happening at different times of the Rabi 
\vor{period}. The 
\vor{loci of} these events and their time spacing can be understood when overlapping the $s=0$ isocontent (white solid) lines 
with the vortex dipole in the relative phase \vor{profile, as shown in}  Fig.~\ref{FIG_asymmetric_case}c.
It is where the isophase lines are tangent to the isocontent \lmd{[or, in other terms, $\bm{\nabla}\varphi_{LU} \perp \hat{\bm{s}}_{orbit}$, with $\hat{\bm{s}}_{orbit}$ the local versor of the $s(x,y)=0$ line]} that the \vor{V-AV} pairs are generated and annihilated. The \vor{created} antivortex is accelerating to an infinite speed and \vor{goes to infinite} distance before coming back from the opposite 
direction \vor{of the cloud} to \vor{participate in} the annihilation process. However, the \vor{photonic} vortex cores 
are only visible inside the packet density, which is fading down with distance and time. 
\rev{The time interval between the pair creation and annihilation events is evaluated as $1/4~T_R$ (if $S$ and distance $d_{LU}$ are kept constant) by counting the isophase lines between them. From the previous  Fig.~\ref{FIG_table_cases}a it can be inferred that, in different situations, not only the orbits but also the timing of the V-AV events is changed, depending on the isocontent and isophase lines' relative positions and their intersections. At smaller separations $d_{LU}$, the time interval of the V-AV pairs existence is reduced.}\\

\begin{figure}[htbp]
  \centering \includegraphics[width=8cm]{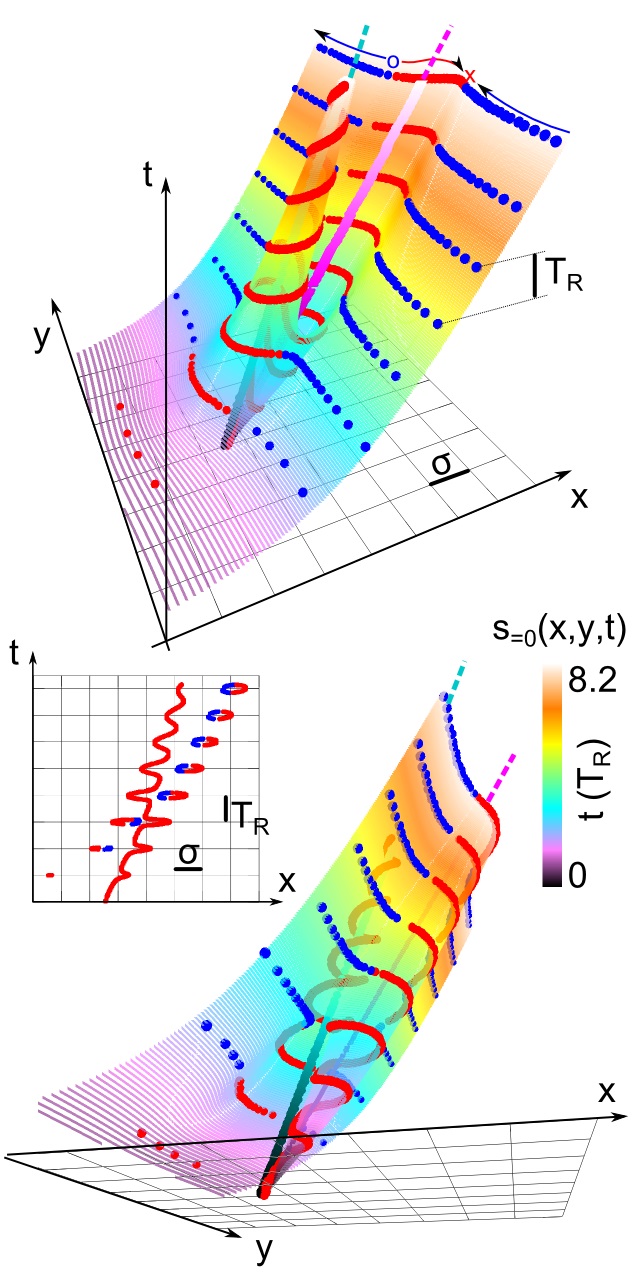}
  \linespread{1.0} \protect\protect
  \caption{\textbf{Topology of the $s=0$ isocontent lines as an $xyt$ surface.}
    \textbf{a, b}, The two plots depict 
    two projections of
    the evolution of the isocontent lines $s(x,y)=0$ as a function of time (vertical axis). The two views are from the top and bottom of the $xy$--plane, for the sake of clarity.
    The cone surface visible on the 
    \lmd{bottom} panel is emerging from a point-like source which represents the initially overlapped vortex core in the UP and LP fields (purple and green lines, respectively), splitting in time and enlarging this cone. The cone visible in the 
    \lmd{top} panel represents the isocontent line wrapping around the slower \vor{moving vortex} core in the LP mode. The main charge (red dots) is 
    \vor{initially} circling around the \lmd{bottom} cone 
    and then passing to the \lmd{top} one, 
    with opposite spiraling directions. The opposite charges (red and blue dots for V and AV, 
    respectively) which are cyclically created and annihilated (blue circle and red cross in the top panel) 
    move on the flatter surfaces 
    following the direction of the arrows.
    The dynamics are represented with respect to the absolute frame of reference, ({\it i.e.}, their tilt to the right side is due to the vortex packets velocity). The dashed lines are a continuation to the LP and UP cores straight trajectories, added for clarity.
    \dom{The inset represents an $xt$ projection, where the V--AV lines appear as closed loops.}
}
\label{FIG_isocontent_surface}
\end{figure}
%

\noindent \textbf{2.8 Topology of the isocontent surface.}
The dynamics of the vortex cores within the evolution of the 
\vor{population imbalance} and \vor{relative} phase of the normal modes represent an exemplary case of complex light, involving singular beams, group velocities, Rabi oscillations and the Bloch sphere mapping. The complexity of the topology can be represented as surfaces and lines in a 3D space when considering the $xyt$ domain. In Fig.~\ref{FIG_isocontent_surface}, we present the model isocontent surface $s_{=0}(x,y,t)$, from different angles of perspective for clarity. 
In the bottom panel it is evident that at the initial time, the LP and UP \vor{vortex} cores are overlapped and start to separate as straight lines (green and purple, respectively) with their relative linear movement. The change with time of the global content \vor{imbalance} $S$ gives rise to the diverging and converging cone wrapped around such vortex lines generator, with the inversion of one cone into the other at the merging of the two surfaces. The photonic main vortex core is shown (red dots) spiraling around the two opposite cones at different time intervals. On the other hand, the secondary vortex-antivortex pairs (red and blue dots, respectively) are present on the open branch of the $s=0$ surface.
\dom{Each pair belongs to what in reality are unitary and continuous vortex lines in the $xyt$ domain. They appear as a pair creation and annihilation events when intersecting such lines with the time sequence of space-like foliations~\cite{freund_critical_2001,berry_much_1998} ({\it i.e.,} in the $xy$ surfaces), see also the loops in the inset in Fig.~\ref{FIG_isocontent_surface} for a different projection.}
The almost horizontal slope of the trajectories and the rarefaction of the dots, 
\dom{seen at the positions of high flatness of the surface,
implies their very fast speed in the $xy$--plane,
to be contrasted with the slow velocity
in the high curvature regions of the cones.}
\dom{It should be noted that none of the vortex lines represent a null density of the full-wavefunction, that is, $|\psi_{L}|^2+|\psi_{U}|^2$ is nowhere zero.
Each vortex core is a phase singularity and a zero density in the associated field (C or X or any other).}
This is only one example of the topological complexity at reach with simple fundamental blocks of singular $\text{LG}$ beams when powered into a platform of strongly coupled fields.
It represents a basic scheme for understanding sophisticated topological concepts and more advanced ways of dynamically shaping complex light, which also involves the oscillations of OAM.\\

\begin{figure}[htbp]
  \centering \includegraphics[width=8cm]{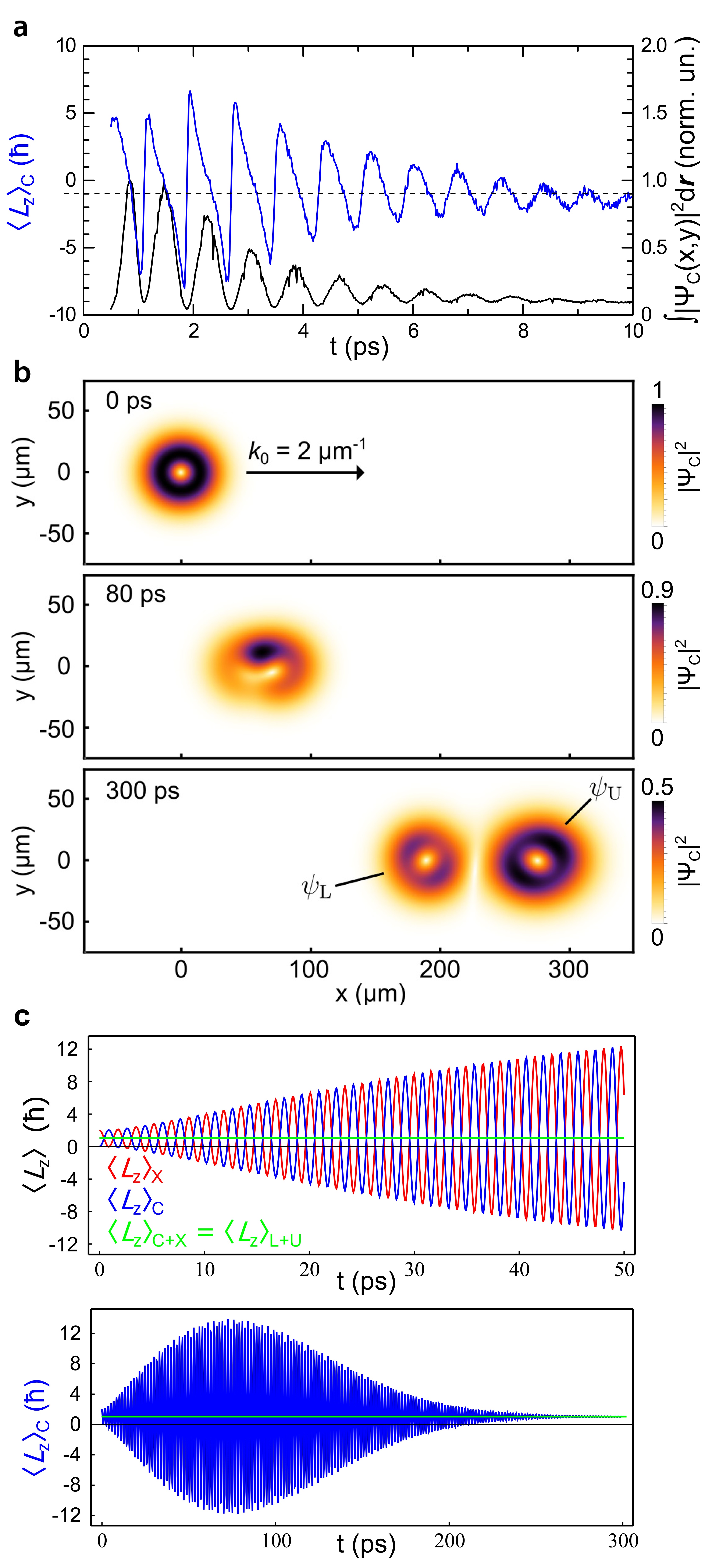}
  \linespread{1.0} \protect\protect
  \caption{\textbf{Mean OAM evolution.}
    \textbf{a}, The average OAM $\langle L_z\rangle_C$ in the experimental photon field as a function of time, left axis, and total photon population (area integrated density), right axis. The OAM is computed with respect to the fixed pole represented by the initial vortex core position.
    \textbf{b}, \vor{Moving Rabi-oscillating vortex modelisation showing the photon density $|\psi_C|^2$ at three different times of the vortex motion ($t=0$~ps, 80~ps, and 300 ps) based on the coupled Schr\"{o}dinger equations. At later times, the vortices in the LP and UP components get completely separated, losing the overlap.
    \textbf{c}. Top: OAM expectation value of the different fields (C in red, X in blue) versus time. The net OAM (green line) is conserved. Bottom: OAM for the photonic field only, computed for larger times. It evolves with the change of the spatial overlap between the LP and UP shown in {\bf b}.}
\rev{The initial OAM per particle is $-\hbar$ in the experimental case and $+\hbar$ in the model case.}
    \lmd{See also Visualization 3.} 
}
\label{FIG_moving_oam}
\end{figure}

\noindent \textbf{2.9 OAM oscillations.}
\lmd{Finally,} \vor{since the dynamics of photonic-mode projection of the polariton full wavefunction features vortex core self-oscillations and cyclic two- or three-charges events, as described above, the photoluminescence from the microcavity possesses a periodically changing topology ({\it i.e.}, a combination of angular momentum oscillations with the linear momentum of the moving vortex). In this section,}
\lmd{we present the oscillations which are seen in the average OAM 
of the emitted 
light $\langle L_{z}\rangle_C$.
\rev{In Fig.~\ref{FIG_moving_oam}a, we report the experimental case (the same as in Fig.~\ref{FIG_exp_moving} and Fig.~\ref{FIG_early_times}a,b with an initial negative vortex charge) while in Fig.~\ref{FIG_moving_oam}b,c the numerical case (starting with a positive vortex and in absence of damping) is shown.}
We have used a fixed point, which is the initial \vor{photonic vortex} core position, as a pole with respect to which we compute the OAM.
In the case of a symmetric vortex beam, the OAM would consist of the iOAM \vor{(the initially imprinted vortex charge)} only, and would remain quantized and constant, independently from the pole with respect to which it is computed. \vor{Here,} the cyclical movement of the \vor{vortex} core position outside and inside the moving packet is instead associated with the changes in the total OAM due to its varying extrinsic component.
In the case of a single vortex, 
the total OAM with respect to a specific point $o$ (such as the initial position) can be written as $\bm{L}_o = \bm{L}_{i} + \bm{r}_{oi} \times \langle\bm{k}\rangle$, where $\bm{L}_{i}$ is the quantized and constant iOAM computed with respect to the instantaneous vortex core, $\bm{r}_{oi}$ is the position vector of the core with respect to the fixed point, and $\langle\bm{k}\rangle$ the mean transverse linear momentum (per particle). For a standing symmetric vortex beam, $\bm{r}_{oi}$ is constant in time and $\langle\bm{k}\rangle$ is null due to symmetry. 
However, here they both change due to the spiraling and the translation velocity. While the spiraling alone would lead to the oscillating OAM, which can only be reduced below the quantized charge value, the joint effect of the increasing distance with the asymmetric reshaping makes the $\langle L_z \rangle\vor{_C}$ oscillations sweep over several values of the quantized unit $\hbar$, see Fig.~\ref{FIG_moving_oam}a,c.
The total OAM at the same time is conserved if one considers the full wavefunction, composed of both the photon and exciton fields (or, otherwise, upper and lower \vor{polariton} modes, see the straight horizontal line in Fig.~\ref{FIG_moving_oam}c).} 

\lmd{The oscillating OAM observed in the emitted photonic component of the moving Rabi\vor{-oscillating} vortex is another exemplary case of the time-varying OAM~\cite{rego_generation_2019}. It is interesting to note that the OAM oscillations are initially growing and then decreasing due to the population imbalance, see Fig.~\ref{FIG_moving_oam}a, similarly to what was previously observed in the FBB case~\cite{dominici_full-bloch_2021}. 
Here, due to the combination of OAM oscillations with the linear momentum, we observe in the emitted light $\langle L_z \rangle_C$ going over $5\hbar$ during the dynamics, converging with time to the initially imprinted unitary charge \rev{($-\hbar$ in the experimental case and $+\hbar$ in the model case)} due to the differential decay.
As in any interference effect, the modulation is in fact maximum when the two fields have equal amplitudes, which is reached when $S=0$. However, in the present case, there is also the effect of the relative movement which is changing the spatial overlap between the two packets (Fig.~\ref{FIG_moving_oam}b), hence altering the modulation strength and oscillations in both the bare C,X fields density and OAM.} 
Indeed, in the theoretical ideal case of no decay which is shown in Fig.~\ref{FIG_moving_oam}c, one sees amplification of OAM oscillations up to $\pm12\hbar$, which then decay because of the gradual separation of the UP and LP wavepackets. It should be underlined that such amplification from $\hbar$ to over $12\hbar$ happens without any nonlinearity at play, but only because of the interplay of the Rabi oscillations and the differential group velocities of the two normal modes of the system.\\\\


\noindent {\large \textbf{3. Conclusions and Perspectives}}\\
We have investigated the concepts at the basis of full Bloch and full Poincar\'{e} beams by implementing ultrafast moving vortices into a planar two-component fluid of  polaritons.
The experimental results consist of an \vor{overall moving} photon vortex core, 
spiraling 
\vor{along} diverging and converging circles, and additional secondary vortex-antivortex pair-creation and -annihilation events \vor{cyclically taking place during the fluid's Rabi oscillations}.
\vor{This behavior leads to the periodically changing  angular momentum directly observable in the light emitted from the device, oscillating in the presented case with an amplitude exceeding $5\hbar$.} The modeling of the dynamics confirms the nature of the observed phenomenology as relying on the overlap of the two polariton modes bearing a spatially-varying 
\vor{population imbalance} and a geometrical phase difference, in the shape of a vortex dipole, \vor{both evolving in time due to their different group velocity, the differential decay of the modes and the Rabi oscillations,} 
hence interfering in such a structured way.
The mapping of the \vor{polariton} Bloch sphere surface to the real space is not homeomorphic. 
\dom{It has been shown in the previous FBB case how such bijective topological mapping}
could subsist and
coincide with a conformal stereographic projection between the 
sphere and the 
plane in the case of a superposition of \vor{concentric} $\text{LG}_{00}$ and $\text{LG}_{01}$ \vor{beams} 
of the same 
width
~\cite{dominici_full-bloch_2021,beckley_full_2010}. In the present case, that special situation is 
\vor{destroyed} by the fact that the wavepackets are moving with different velocities so that the centres of the two UP/LP $\text{LG}_{01}$ \vor{profiles} 
\vor{become} displaced. 
\dom{On the contrary to the homeomorphism, the current situation leads to the degeneracy of multiple real space points which are mapped to the very same polariton state, {\it i.e.}, point on the sphere. Such a situation is indeed physically observed by tracking a secondary vortex-antivortex pair, created and subsequently annihilated, in the emitted light, where each cores of V and AV, being a zero of the photon field, represents the pure excitonic state.}
More complex situations are within reach with the case of multiple charge imprinting or vortex lattices, or by adding the polarization control thus making dynamical full Poincar\'{e} beams.
\rev{Using for example two opposite unitary charges ($\text{LG}_{0+1}$ and $\text{LG}_{0-1}$) in the UP and LP modes would lead to the same isocontent morphologies but with the relative phase being that of a double charge rather than a dipole. A vortex core would be seen starting to orbit in the photon (or exciton) field, this time around both the LP and UP vortex cores, and an antivortex slowly coming into the packet to replace it at long times, after a transition time of the two orbiting each other.}
Implications touching the concepts of topology and Berry curvature will be shown in further works, while the possible use for optical tweezers, particles sorting, data encoding and interferometry gyroscopes remain to be investigated.\\



\noindent {\large \textbf{Methods}}\\
\noindent \textbf{Experimental methods}
\rev{The polariton device used here is fabricated by means of MBE technique and consists in an AlGaAs $2\lambda$ MC with three In$_{0.04}$Ga$_{0.96}$As quantum wells of 8 nm placed at the antinodes of the cavity mode field. The cavity is placed inside two distributed Bragg reflectors made of 21 and 24 pairs of alternated $\lambda/4$ AlAs and GaAs layers. The experiments are performed in a region of the sample clean from defects. The device is kept at a temperature of 10 K inside a closed-loop He cryostat.}
The resonant beam is a 130~fs pulse laser with 80~MHz repetition rate and a 8~nm bandwidth. The central energy is tuned \rev{at approximately 833.5 nm} in order to overlap both the LP and UP branch energies at the selected finite momentum. The photonic beam is passed through a $q$-plate device~\cite{marrucci_optical_2006,karimi_efficient_2009}, in order to obtain a unitary optical vortex which is sent onto the microcavity sample at oblique incidence. The polariton sample modes splitting is 3~nm (5.4~meV) at zero momentum which converts into a Rabi period of $T_R =0.780$~ps
\rev{while the $T_R$ at the finite momentum of the experiments is slightly lesser.} 
Once imprinted, the polariton vortex is free to move and evolve. The lifetime of the normal modes is $\tau_{L}=1/\gamma_{L}\sim 10$~ps and $\tau_{U}=1/\gamma_{U}\sim 2$~ps for the lower and upper modes, respectively.
An ultrafast time resolved detection scheme is based on the off-axis digital holography, where the emission from the sample plane is focused on an imaging camera together with a reference beam. The reference pulse is derived by the resonant laser beam and is not focused but passed through a small iris, so that it becomes a homogeneous and flat front when arriving on the camera. The digital fast Fourier transform (FFT) filtering is applied to the interferograms and this procedure allows to reconstruct the amplitude and phase of the emitted photonic wavefunction. The FFT filtering can be applied directly in the laboratory in order to aid for the setting of the measurements.
For additional details on both the polariton sample and digital holography technique see Refs~\cite{Dominici2014,colas_polarization_2015,dominici_full-bloch_2021} and descriptions therein.

\vspace{0.2cm}

\noindent \textbf{Numerical  methods}
The 
model is set by use of \vor{the two} $\text{LG}_{01}$ 
\vor{packets} moving with different velocities, 
\vor{representing} the upper and lower polariton fields. The bare exciton and photon fields are retrieved as linear combination of the normal modes.
We write the spatial profiles of the fields in complex polar coordinates $(r,\phi)$, where $r=|x+iy|$ and $\phi = \arg(x+iy)$. The fields are written as\\

$
\psi_{U,L} = \\
 A_{U,L} e^{-|\bm{r}-\bm{r}_{U,L}|^2/2\sigma^2} \frac{|\bm{r}-\bm{r}_{U,L}|}{\sigma} e^{i \arg(\bm{r}-\bm{r}_{U,L})} 
e^{i k_0 x} e^{(i\omega_{U,L}-\gamma_{U,L})t}\\
$

The mo\vor{tion} 
of the two packets is \vor{simulated} 
by updating the \vor{distance to the} centres of the beams with their group velocities $\bm{r}_{U,L}(t) = \bm{v}_{U,L}t$, assumed to be along the $x$-axis.
This simple model gives the same results than would be derived with a full coupled Schr\"{o}dinger equations (cSE) model, apart from neglecting the dispersion and diffraction effects and the initial transient from the incoming photon pulse being injected into the polariton components. We are not interested in the sub-ps transient and the beams are wide enough in order to neglect any dispersive/diffraction effects in the 10 ps time scale range. 
Assuming a zero detuning between the cavity photon to exciton mode, the bare modes are hence obtained as
$\psi_{C,X} = \vor{(}\psi_{U} \pm \psi_{L}\vor{)/\sqrt{2}}$.
\lmd{For completeness, the cSE model has been used to verify the behavior of the moving Rabi-oscillating vortex, of the same kind as described in \cite{dominici_full-bloch_2021}, and adding the initial central momentum \vor{$k_0$} to the photonic $\text{LG}_{01}$ resonant pulse.}\\


\noindent \textbf{Funding.} 
Russian Foundation for Basic Research (joint with CNR, Project No. 20–52–7816); Ministero dell’Istruzione,
dell’Università e della Ricerca (PRIN project InPhoPol); Ministero dell’Istruzione, dell’Università e della Ricerca (CUP:
B83B17000010001, delibera CIPE n. 3449 del 7/08/2017, FISR MIUR CNR, TECNOMED); Regione Puglia (CUP:
B84I18000540002, DGR n. 2117 del 21/11/2018, TecnoMed Puglia); Iran National Science Foundation (97009377).\\

\noindent \textbf{Acknowledgements.} 
We would like to thank Romuald Houdr\'{e} and Alberto Bramati for the microcavity sample, Lorenzo Marrucci and Bruno Piccirillo for the $q$-plate devices, and sir Michael V.~Berry for interesting comments.\\

\noindent \textbf{Disclosures.}
The authors declare no conflicts of interest.\\

\noindent \textbf{Data availability.} 
Data underlying the results presented in this paper are not publicly available at this time but may be obtained from the authors upon reasonable request.\\

\noindent \textbf{Journal Ref.} 
Cite this work as 
\href{https://doi.org/10.1364/OE.438035}{Opt. Express \textbf{29}, 37262 (2021)}.\\

\noindent {\large \textbf{Videos}}\\
\noindent \textbf{Visualization 1.}  
Experimental moving Rabi oscillating vortex seen in the photonic component as in Fig.~\ref{FIG_exp_moving}a and Fig.~\ref{FIG_early_times}a,b. Photonic amplitude and phase in a $60 \times 60~\mu\text{m}^2$ area, with 20 fs time step over a 10 ps time span. The photonic pulse beam is arriving at around $t = 0.8~\text{ps}$. The vortex core positions are marked with a red dot in the amplitude map.\\

\noindent \textbf{Visualization 2.} 
The $\text{LG}$s model for the moving Rabi oscillating vortex, comprising the maps and dynamics
of four relevant quantities.
The four panels show the 
local imbalance $s(\bm{r},t)$ (purple-green map) and the relative phase $\varphi_{LU}(\bm{r},t)$ (black and white map)
on the top row,
as in Fig.~\ref{FIG_table_cases} and Fig.~\ref{FIG_asymmetric_case},
together with the photon and exciton densities $|\psi_\mathrm{C,X}(\bm{r},t)|$
(blue and red maps, respectively) in the second row.
The positions of the translating UP/LP cores are corresponding to the two poles of the vortex dipole in the relative phase map.
The more complex motion of the spiraling and branching photon and exciton cores, seen as holes in their respective fields, are tracked by the crossing points  
between the $s=0$ isocontent line (white lines)
and the $\varphi_{LU} =0$ and $\varphi_{LU} = \pi$ 
isophase lines (black circles).\\

\noindent \textbf{Visualization 3.} 
Computational moving vortex and OAM dynamics obtained by a cSE model in absence of damping, as represented in Fig.~\ref{FIG_moving_oam}b,c. Both the photon
and exciton components are shown, during the vortex packet
propagation. The bottom panel shows the OAM oscillations,
increasing with the initial packets’ separation and then
lowering down due to decrease of the packets’ overlap at later
times. The simulation allows also to appreciate the UP mode
(leading vortex) and LP mode (trailing vortex), moving with different group velocities.



%





\end{document}